\documentclass[11pt]{llncs}
\usepackage{amssymb,amsfonts}
\usepackage{graphicx}
\newcommand{\trace}{\mathop{\rm Tr}\nolimits}

\newcommand{\diag}{\mathop{\rm Diag}\nolimits}

\newcommand{\supp}{\mathop{\rm supp}\nolimits}

\newcommand{\twomat}[4]{\left(\begin{array}{cc}#1&#2\\#3&#4\end{array}\right)}

\newcommand{\cH}{{\mathcal H}} 
\newcommand{\cE}{{\mathcal E}} 
\newcommand{\cT}{{\mathcal T}} 

\DeclareRobustCommand\openone{\leavevmode\hbox{\small1\normalsize\kern-.33em1}}
\newcommand{\id}{\mathrm{\openone}}

\newcommand{\be}{\begin{equation}}
\newcommand{\ee}{\end{equation}}
\newcommand{\bea}{\begin{eqnarray}}
\newcommand{\eea}{\end{eqnarray}}
\newcommand{\beas}{\begin{eqnarray*}}
\newcommand{\eeas}{\end{eqnarray*}}
\newcount\minute
\newcount\hour
\def\currenttime{%
    \minute\time
    \hour\minute
    \divide\hour60
    \the\hour:\multiply\hour60\advance\minute-\hour\the\minute}

\begin{document}
\title{Telescopic Relative Entropy}
\author{Koenraad M.R.\ Audenaert}
\institute{Department of Mathematics,\\
Royal Holloway, University of London,\\
Egham TW20 0EX, United Kingdom}

\maketitle
\begin{abstract}
We introduce the telescopic relative entropy (TRE), which is
a new regularisation of the relative entropy related to smoothing,
to overcome the problem that the relative entropy between
pure states is either zero or infinity and therefore useless as a distance measure in this case.
We study basic properties of this quantity, and find interesting relationships between the TRE and the
trace norm distance. We then exploit the same techniques to obtain a new and shorter proof of a lower bound
on the relative Renyi entropies in terms of the trace norm distance, $\trace\rho^{1-p}\sigma^p \ge 1-||\rho-\sigma||_1/2$. 
\end{abstract}

\section{Introduction\label{sec:intro}}
The quantum relative entropy between two quantum states $\rho$ and $\sigma$,
$S(\rho||\sigma)=\trace\rho(\log\rho-\log\sigma)$, is a non-commutative generalisation
of the Kullback-Leibler distance between probability distributions.
Because of its strong mathematical connections with von Neumann entropy, and its interpretation as
an optimal asymptotic error rate in quantum hypothesis testing (in the context of Stein's lemma)
relative entropy is widely used as a (non-symmetric) distance measure
between states \cite{ohya_petz}.

One of its drawbacks, however, is that for non-faithful (rank-deficient) states the relative entropy can be
infinite. More precisely, the relative entropy is infinite when
there exists a pure state $\psi$ such that $\langle\psi|\sigma|\psi\rangle$
is zero while $\langle\psi|\rho|\psi\rangle$ is not.
In particular, relative entropy is useless as a distance measure between pure states,
since it is infinite for pure $\rho$ and $\sigma$, unless
$\rho$ and $\sigma$ are exactly equal (in which case it always gives $0$).

There are various possibilities to overcome this deficiency.
In \cite{lendi}, Lendi, Farhadmotamed and van Wonderen proposed a \textit{regularised relative entropy} as
$$
R(\rho||\sigma)=c_d\,\, S\left(\frac{\rho+\id_d}{1+d} \Bigg|\Bigg|\frac{\sigma+\id_d}{1+d}\right),
$$
where $d$ is the dimension, and $c_d$ is a normalisation constant. This only works for finite-dimensional states.

Another possibility, also useful for infinite dimensional states,
is to apply a smoothing process. One can define the \textit{smooth relative entropy}
between states $\rho$ and $\tau$ as
the infimum of the ordinary relative entropy between $\rho$ and another state $\tau$, where $\tau$
is constrained to be $\epsilon$-close to $\sigma$ in trace norm distance:
$$
S_\epsilon(\rho||\sigma) = \inf_{\tau} \left\{S(\rho||\tau):
\tau\ge0, \trace\tau\le1,||\tau-\sigma||_1\le\epsilon\right\}.
$$
This form of smoothing has already been applied to Renyi entropies and max-relative entropy \cite{nila,renner},
giving rise to a quantity with an operational interpretation,
but it could equally well be applied to ordinary relative entropy.

In the case of the ordinary relative entropy there is a simple canonical choice for $\sigma$ that achieves
the same purpose of regularisation but without having to find the exact minimiser.
Namely, we can take that $\tau$ that is collinear with $\rho$ and $\sigma$; i.e.
$\tau=a\rho+(1-a)\sigma$ (with $a=\epsilon/||\rho-\sigma||_1$).

By operator monotonicity of the logarithm, we have
$$
\log(\tau)=\log(a\rho+(1-a)\sigma)\ge \log(a\rho),
$$
and, therefore,
\beas
S(\rho||\tau)&=&\trace\rho(\log\rho-\log\tau) \\
&\le& \trace\rho(\log\rho-\log(a\rho)) \\
&=& -\log a.
\eeas
Thus, $S(\rho||\tau)$ is bounded above by $-\log a$, which is finite for $0<a<1$.
It therefore makes perfect sense to normalise $S(\rho||\tau)$ by dividing it by $-\log a$, producing a quantity
that is always between $0$ and $1$.

These observations led us to define what we call the
\textit{telescopic relative entropy} (TRE), a particular regularisation of the ordinary
relative entropy that is also defined in Hilbert spaces of infinite dimension:
\begin{definition}
For fixed $a\in(0,1)$, the $a$-telescopic relative entropy between states $\rho$ and $\sigma$ is given by
\be
S_a(\rho||\sigma) := \frac{1}{-\log(a)} \,\,S(\rho||a\rho+(1-a)\sigma).
\ee
Furthermore, we define
\bea
S_0(\rho||\sigma) &:=& \lim_{a\to 0}S_a(\rho||\sigma) \\
S_1(\rho||\sigma) &:=& \lim_{a\to 1}S_a(\rho||\sigma).
\eea
We'll show below that these limits exist.
\end{definition}
The origin of the name is that the operation $\sigma\mapsto a\rho+(1-a)\sigma$ acts like a `telescope'
with `magnification factor' $1/(1-a)$, bringing the state $\sigma$ closer to the `vantage point' $\rho$
and bringing observed pairs of states $\sigma_i$ closer to each other.

The purpose of this paper is to initiate the study of this quantity.
The telescoping operation $\sigma\mapsto a\rho+(1-a)\sigma$
and subsequent scaling of the relative entropy by $1/(-\log a)$ may seem like a fairly innocuous operation,
but has a number of far-reaching and sometimes unexpected consequences.
Because of the linearity of the telescoping operation, the TRE inherits most of the desirable properties of the
ordinary relative entropy.
However, a host of additional relations in the form of sharp inequalities may be derived that in the case
of the ordinary relative entropy simply make no sense, because the constants appearing in the inequality
would be infinite.
At the end of this paper, we briefly consider the telescoping operation in the context of the relative Renyi entropies.
We exploit the same techniques used for the TRE to obtain a new and shorter proof of a lower bound
on the relative Renyi entropies in terms of the trace norm distance, $\trace\rho^{1-p}\sigma^p \ge 1-||\rho-\sigma||_1/2$
\cite{ka}.
\section{Preliminaries\label{sec:pre}}
For any self-adjoint operator $X$ on a Hilbert space $\cH$,
we denote by $\supp X$ the support of $X$, i.e.\
the subspace of $\cH$ which is the orthogonal complement of $\ker X$, the kernel of $X$.
The projector on the support of $X$ will be denoted by $\{X\}$.
We denote by $P_X$ the orthogonal projector from $\cH$ onto $\supp X$,
so that $P_X^*$ is the injection of $\supp X$ back into $\cH$. Thus $P_X^* P_X = \{X\}$.
The \textit{compression of $A$ to the support of $X$},
which we'll denote by $A|_X$, is the operator with domain
$\supp X$ given by
\beas
A|_X &=& P_X A P_X^*.
\eeas
By definition, for any positive operator $X\ge0$, we have $X|_X>0$, strictly.

Two quantum states are mutually orthogonal, denoted $\rho\perp\sigma$, iff $\trace\rho\sigma=0$.

For any self-adjoint operator $X$, $X_+$ will denote the positive part $X_+ = (X+|X|)/2$.
It features in an expression
for the trace norm distance between states:
\be
T(\rho,\sigma) := \frac{1}{2}||\rho-\sigma||_1 = \trace(\rho-\sigma)_+.
\ee
The trace of the positive part has the variational characterisation
$\trace X_+ = \max_P \trace XP$, where the maximisation is over all self-adjoint projectors.
Hence, for all such projectors $P$, $\trace XP\le \trace X_+$.

The Pinsker bound is a
lower bound on the ordinary relative entropy in terms of trace norm distance, \cite{ohya_petz}.
\be
S(\rho||\sigma) \ge 2T(\rho,\sigma)^2.
\label{eq:pinsker}
\ee
No upper bound in terms of the trace norm distance is possible, because the relative entropy can be infinite.

We will also need the following integral representation of the logarithm: for $x>0$, we have
\be
\log x = \int_0^\infty ds \left(\frac{1}{1+s}-\frac{1}{x+s}\right).\label{eq:intlog}
\ee
This immediately provides an integral representation for the telescopic relative entropy:
\bea
\lefteqn{S_a(\rho||\sigma)} \nonumber \\
&=& \frac{1}{\log a}\,\,\int_0^\infty ds\,
\trace \rho [(\rho+s)^{-1}-(a\rho+(1-a)\sigma+s)^{-1}] \label{eq:int1a} \\
&=& \frac{1}{\log a}\,\,\int_0^\infty ds\,
\trace \rho (\rho+s)^{-1}\,\, (1-a)(\sigma-\rho)\,\, (a\rho+(1-a)\sigma+s)^{-1}. \label{eq:int2a}
\eea

Another integral we will encounter is $\int_0^\infty ds \,\,\,x/(x+s)^2$.
For $x=0$, the integral obviously gives $0$. For $x>0$ it gives $1$.
Hence
\be
\int_0^\infty ds\, (\rho+s)^{-1}\,\rho \, (\rho+s)^{-1} = \{\rho\}.\label{eq:intproj}
\ee

From integral representation (\ref{eq:intlog})
we get an expression for the Fr\'echet derivative of the matrix logarithm:
$$
\frac{d}{dt}\Bigg|_{t=0} \log(A+t\Delta)
=\int_0^\infty ds\,\,(A+s)^{-1} \Delta (A+s)^{-1}.
$$
It will be useful to introduce the following linear map, for $A\ge0$:
\be
\cT_A(\Delta) = \int_0^\infty ds\,\,(A+s)^{-1} \Delta (A+s)^{-1}.
\ee
Thus
\be
\frac{d}{dt}\Bigg|_{t=0} \log(A+t\Delta) = \cT_A(\Delta).
\ee
It's easy to check that for $A\ge0$, $\cT_A(A) = \{A\}$.
Thus, for $A>0$, we have $\cT_A(A)=\id$.

From this integral representation it also follows that, for any self-adjoint $A$, $\cT_A$ preserves the
positive semidefinite order: if $X\le Y$, then $\cT_A(X)\le\cT_A(Y)$.
By cyclicity of the trace, we see that the map $\cT_A$ is self-adjoint:
$\trace B\cT_A(\Delta) = \trace\Delta\cT_A(B)$.
Moreover, the map is positive semi-definite, in the sense that
$\trace\Delta\cT_A(\Delta)$ is positive for any self-adjoint $\Delta$.
This follows from the integral representation
and the fact that for positive $X$ and self-adjoint $Y$, $\trace XYXY = \trace(X^{1/2}YX^{1/2})^2\ge0$.

\section{Basic properties of Telescopic Relative Entropy\label{sec:basic}}
From the discussion in the Introduction, we recall that the value of the telescopic relative entropy
is always between $0$ and $1$, even for non-faithful states.
Furthermore, it inherits many desirable properties from the ordinary relative entropy:
positivity, the fact that it is only zero when $\rho$ and $\tau$ are equal (provided $a>0$),
joint convexity in its arguments, and monotonicity under CPT maps.

As we do not restrict the arguments of the telescopic relative entropy to states,
the definition is also applicable (in a useful way) to non-negative scalars:
\be
S_a(b||c) = \frac{b(\log b - \log(ab+(1-a)c))}{-\log a}.
\ee

\begin{figure}[ht]
\begin{center}
\includegraphics[width=5.5cm]{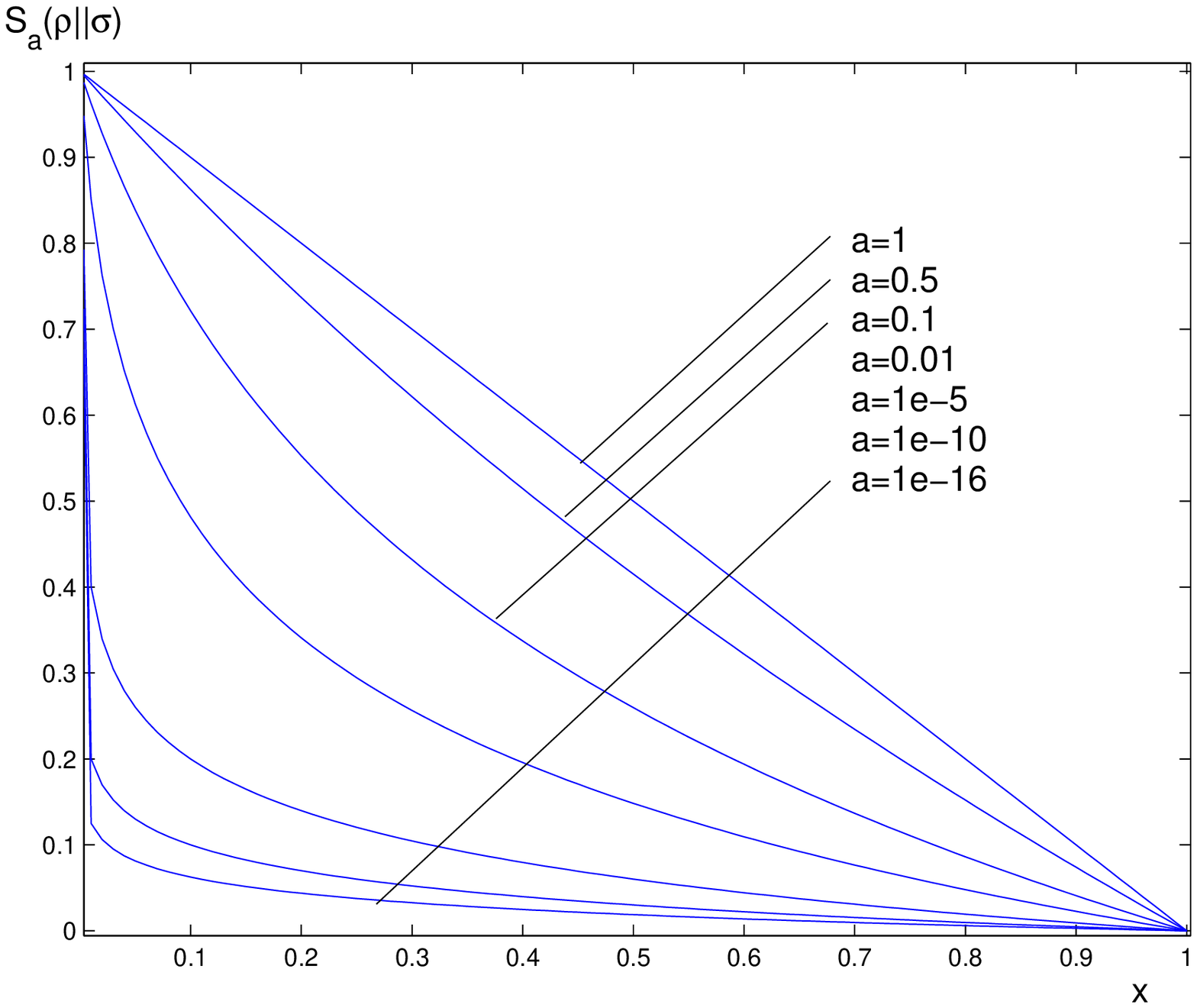}\quad
\includegraphics[width=5.5cm]{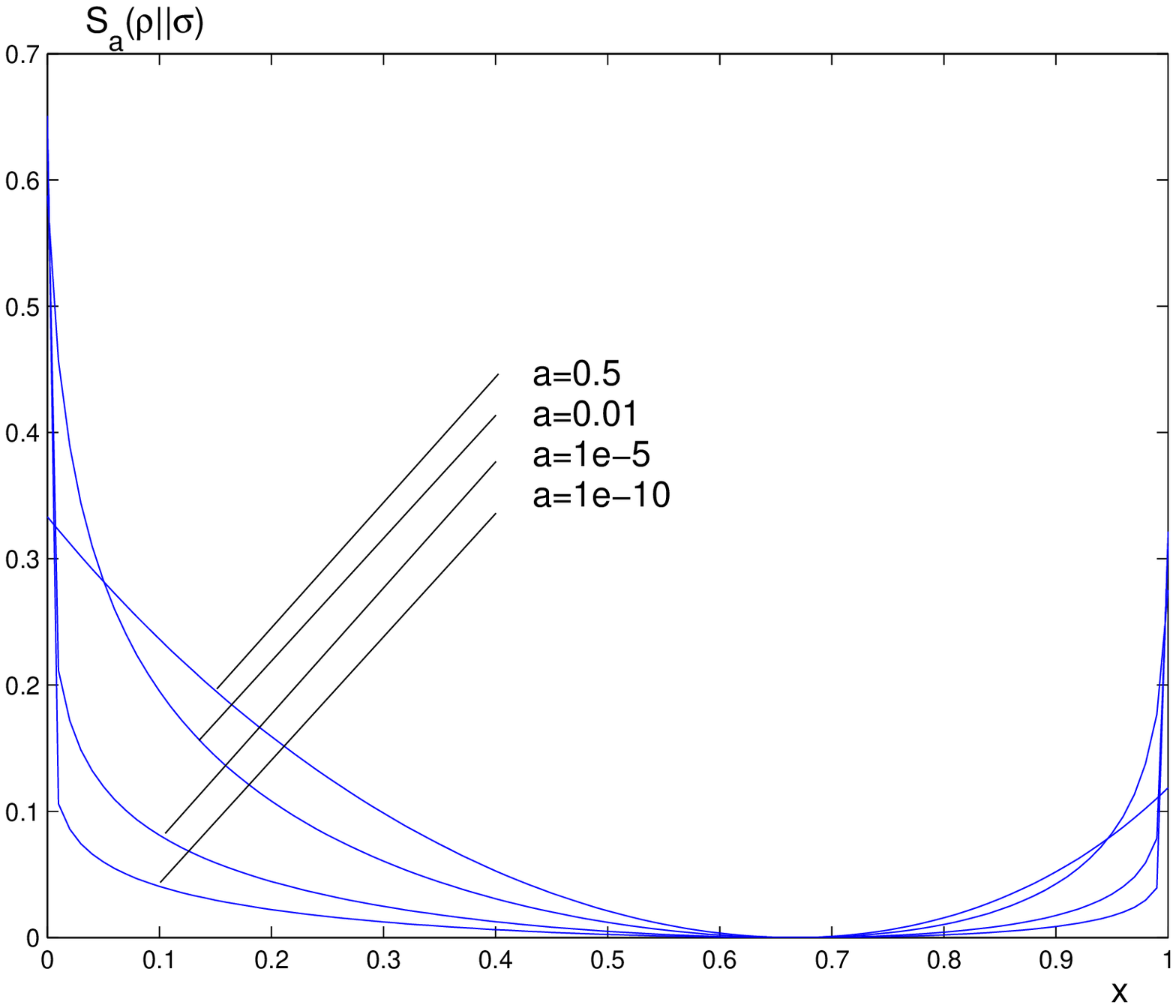}
\caption{(a) Telescopic relative entropy $S_a(\rho||\sigma)$ between state $\rho=|0\rangle\langle0|$
and state $\sigma=x|0\rangle\langle0|+(1-x)|1\rangle\langle1|$, with $x$ ranging from 0 to 1, and
for various values of $a$; (b) same but for $\rho=(2/3)|0\rangle\langle0|+(1/3)|1\rangle\langle1|$.
\label{fig1}
}
\end{center}
\end{figure}

\begin{figure}[ht]
\begin{center}
\includegraphics[width=5.5cm]{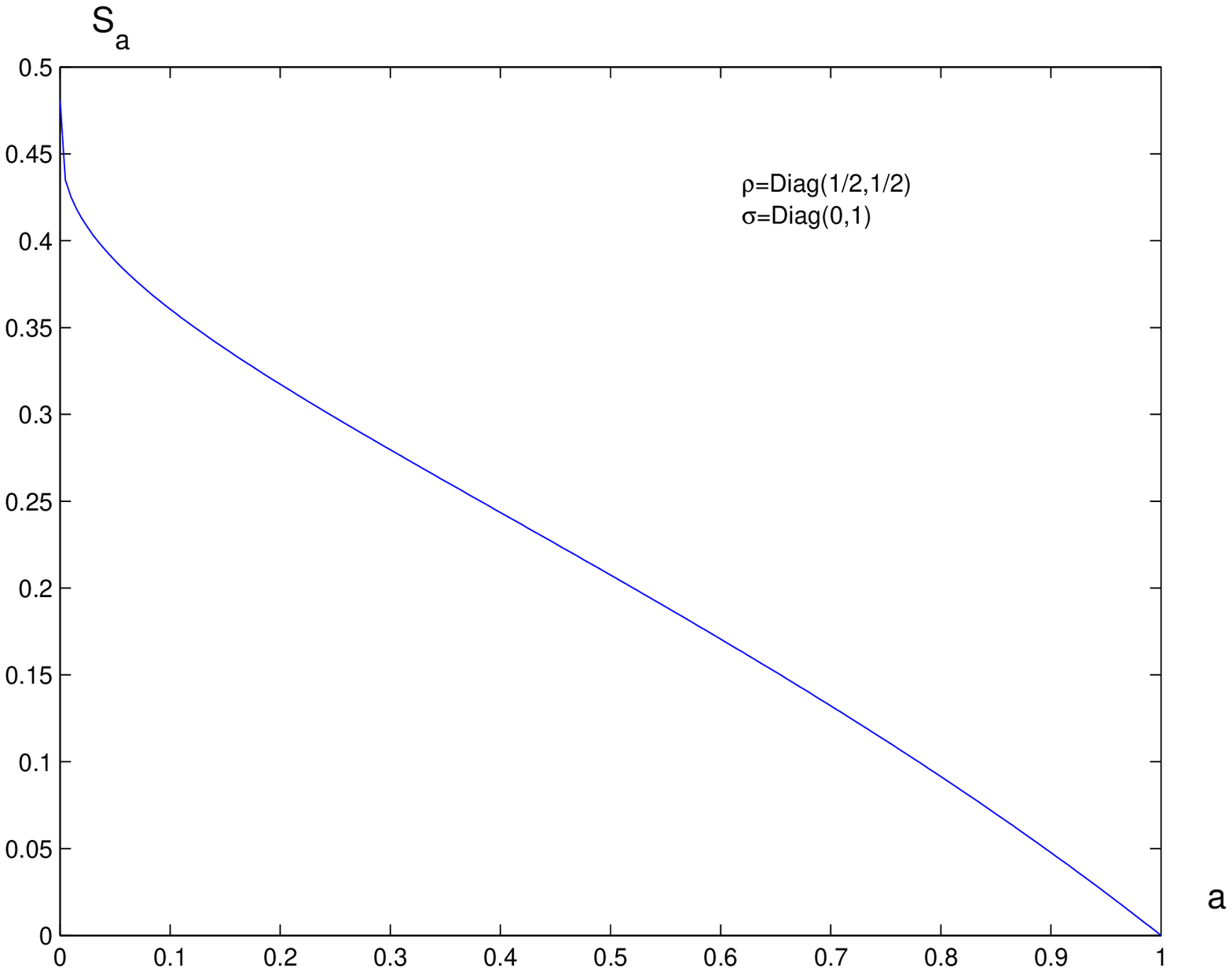}\quad
\includegraphics[width=5.5cm]{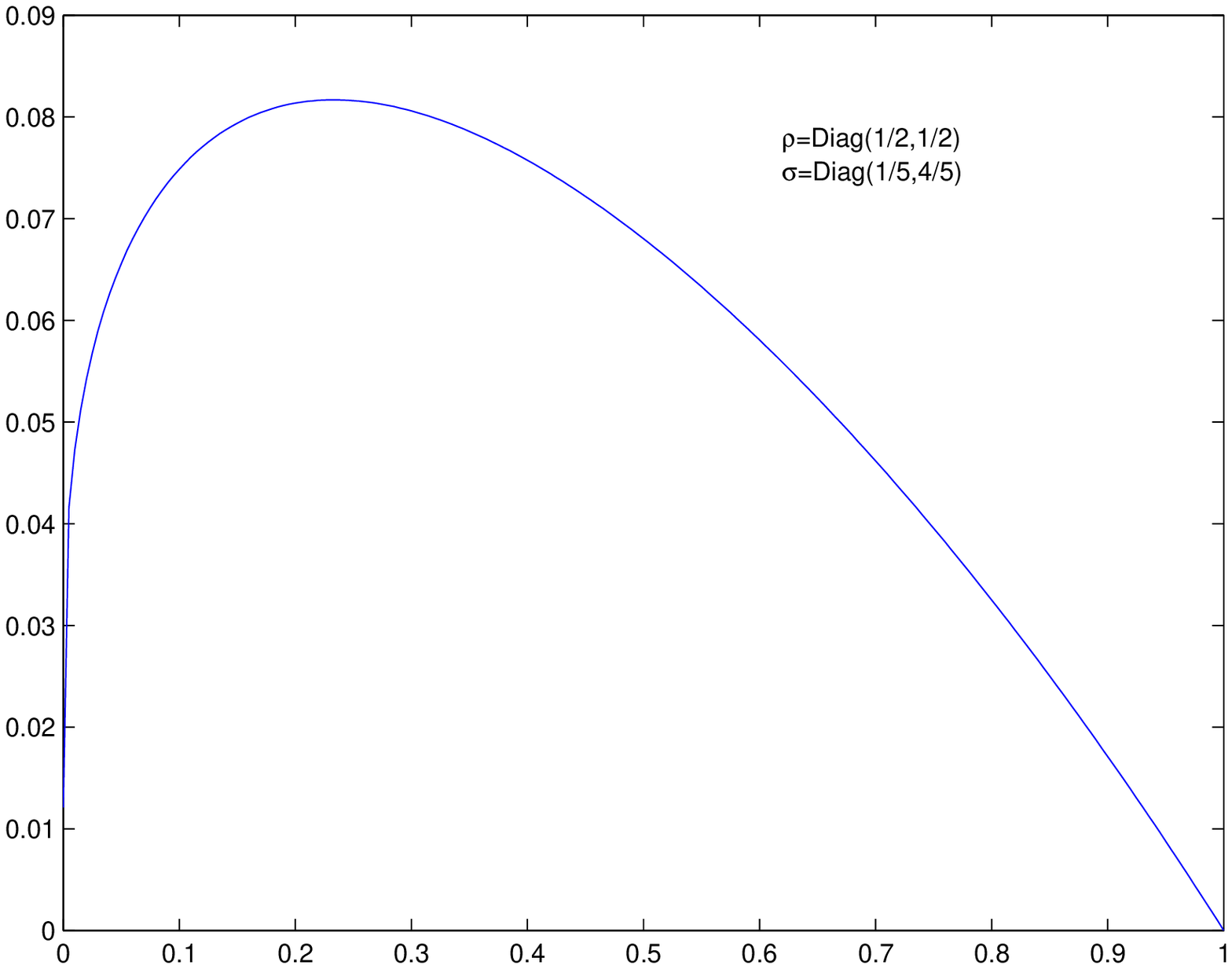}
\caption{(a) Telescopic relative entropy $S_a(\rho||\sigma)$ between state $\rho=\id_2/2$
and state $\sigma=|1\rangle\langle1|$, with $a$ ranging from 0 to 1;
(b) same but for $\sigma=(|0\rangle\langle0|+4|1\rangle\langle1|)/5$.
\label{fig2}
}
\end{center}
\end{figure}

For illustrative purposes, we graph the telescopic relative entropy for a variety of qubit state pairs, in
figures \ref{fig1} and \ref{fig2}.

\subsection{$S_0$ and $S_1$}
One might think that the $1$-telescopic relative entropy would be quite useless, because for $a=1$,
$S(\rho||a\rho+(1-a)\sigma) = S(\rho||\rho) = 0$.
Nevertheless, it is a non-trivial quantity
due to the normalisation by $1/(-\log a)$.
Likewise, one might mistakenly think $S_0$ is essentially the ordinary relative entropy;
it is far from it, and for the same reason. Indeed, for any pair of states with finite ordinary relative entropy,
e.g.\ when both states are faithful, $S_0$ is $0$, due to the normalisation.
The $0$-telescopic relative entropy shows its true colours exactly in those cases
when the ordinary relative entropy yields $+\infty$.

In fact, for $S_0$ and $S_1$ we have the following closed form expressions:
\begin{theorem}\label{th:S01closed}
For any pair of states $\rho$, $\sigma$,
\bea
S_0(\rho||\sigma) &=& 1-\trace \rho\{\sigma\} \\
S_1(\rho||\sigma) &=& 1-\trace \sigma\{\rho\}.
\eea
\end{theorem}
In particular, when $\sigma$ is pure, $S_0(\rho||\sigma)=1-\trace\rho\sigma$, and when $\rho$ is pure,
$S_1(\rho||\sigma)=1-\trace\rho\sigma$.
When $\sigma$ is faithful, $S_0(\rho||\sigma)=0$; when $\rho$ is faithful, $S_1(\rho||\sigma)=0$.

\textbf{Proof.}
Consider first the limit $a\to 1$.
Using de l'H{\^o}pital's rule we find
$$
\lim_{a\to 1} \frac{1-a}{-\log a} = 1.
$$
Hence, by representation (\ref{eq:int2a}),
$$
\lim_{a\to 1} S_a(\rho||\sigma) = -\int_0^\infty ds\,
\trace \rho (\rho+s)^{-1}\,\, (\sigma-\rho)\,\, (\rho+s)^{-1}.
$$
Therefore, from (\ref{eq:intproj}) we get the required
$$
\lim_{a\to 1} S_a(\rho||\sigma) = -\trace (\sigma-\rho)\{\rho\} = 1-\trace \sigma\{\rho\}.
$$

For the limit $a\to 0$ some more work is needed.
Let us w.l.o.g.\ assume that $(\rho+\sigma)/2$ is faithful;
otherwise we take the compression of $\rho$ and $\sigma$ to
the support of $(\rho+\sigma)/2$.
Again we use an integral representation, but in its more basic form (\ref{eq:int1a}).
To calculate the limit $a\to 0$ we apply de l'H{\^o}pital's rule
to the whole expression and get
\beas
\lefteqn{S_0(\rho||\sigma)} \\
&=& \lim_{a\to 0} a \,\, \frac{d}{da}\,\,\int_0^\infty ds\,
\trace \rho [(\rho+s)^{-1}-(a\rho+(1-a)\sigma+s)^{-1}] \\
&=& \lim_{a\to 0} \int_0^\infty ds\,
\trace a\rho (a\rho+(1-a)\sigma+s)^{-1}\,\,(\rho-\sigma)\,\, (a\rho+(1-a)\sigma+s)^{-1} \\
&=& \lim_{a\to 0} \int_0^\infty ds\,
\trace (\rho-\sigma) (a\rho+(1-a)\sigma+s)^{-1}\,\,a\rho\,\, (a\rho+(1-a)\sigma+s)^{-1}.
\eeas
Here, the first factor $a$ comes from the derivative of $\log a$.

Because of our assumption that $(\rho+\sigma)/2$ is faithful, $a\rho+(1-a)\sigma$ is faithful for any $a\in(0,1)$.
Therefore, the integral
$$
\int_0^\infty ds\,
  (a\rho+(1-a)\sigma+s)^{-1}\,\,(a\rho+(1-a)\sigma)\,\, (a\rho+(1-a)\sigma+s)^{-1}
$$
yields the identity operator $\id$. Using this fact, we can rewrite our last expression for $S_0$ as
\beas
\lefteqn{S_0(\rho||\sigma)} \\
&=& \lim_{a\to 0} \trace (\rho-\sigma) [\id -\int_0^\infty ds\, \\
&& \qquad (a\rho+(1-a)\sigma+s)^{-1}\,\,(1-a)\sigma \,\, (a\rho+(1-a)\sigma+s)^{-1}] \\
&=& \trace (\rho-\sigma) [\id -\int_0^\infty ds\, (\sigma+s)^{-1}\,\,\sigma \,\, (\sigma+s)^{-1}] \\
&=& \trace (\rho-\sigma) (\id-\{\sigma\}) \\
&=& 1-\trace\rho\{\sigma\},
\eeas
as required.
\qed

\subsection{Pure states}
From Theorem \ref{th:S01closed} we can derive the equalities
\be
S_0(\rho||\sigma) = S_1(\rho||\sigma) = T(\rho,\sigma)^2,
\ee
for \textit{pure} $\rho$ and $\sigma$.

In fact, when $\rho$ and $\sigma$ are pure, there is a one-to-one relation
between $S_a(\rho||\sigma)$ and $T(\rho,\sigma)$
for any value of $a\in[0,1]$. Although the relation is somewhat complicated, in practice it shows that
$S_a(\rho||\sigma)$ is only slightly bigger than $T(\rho,\sigma)^2$ for $a\in(0,1)$.
\begin{theorem}
Let $\rho,\sigma$ be two pure states with trace norm distance $t=||\rho-\sigma||_1/2$.
Then, for $a\in(0,1)$,
\be
S_a(\rho||\sigma) = \frac{1}{-2\log a}\left(-\log\frac{w}{4}-\frac{1-w/(2a)}{\sqrt{1-w}}
\,\,\log\frac{1+\sqrt{1-w}}{1-\sqrt{1-w}}\right),
\ee
where
\be
w:=4a(1-a)t^2.
\ee
\end{theorem}
\textbf{Proof.}
By a suitable unitary transformation, the problem can be transformed to a two-dimensional one,
with in particular
$$
\rho=\twomat{1}{0}{0}{0},\qquad
\sigma=\twomat{1-t}{\sqrt{t(1-t)}}{\sqrt{t(1-t)}}{t}.
$$
The telescopic relative entropy is then given by
$$
S_a(\rho||\sigma) = \frac{1}{-\log a}\left(-\log\left(a\rho+(1-a)\sigma\right)\right)_{1,1}
$$
and after some basic calculations this reduces to the given formula.
\qed

For example, let $\rho$ and $\sigma$ be two pure two-level states, with the angle between their respective Bloch vectors
equal to $\theta$. Since their trace norm distance is equal to $t=|\sin(\theta/2)|$,
we have $w=2a(1-a)(1-\cos\theta)$.

\section{Comparison to trace norm distance\label{sec:tracenorm}}
In this section,
we provide bounds on the telescopic relative entropy in terms of the trace norm distance.

It's very easy to derive a lower bound
from the  Pinsker lower bound on the ordinary relative entropy (\ref{eq:pinsker}).
\begin{theorem}
For two quantum states $\rho,\sigma$,
\be
S_a(\rho||\sigma) \ge \frac{(1-a)^2}{-\log(a)}\,\, 2\, T(\rho,\sigma)^2.\label{eq:SaHOT}
\ee
\end{theorem}
\textbf{Proof.}
Noting that $T(\rho,\tau) = (1-a) T(\rho,\sigma)$, this is a trivial consequence of the bound
$
S(\rho||\tau) \ge 2\,T(\rho,\tau)^2.
$
\qed

While there is no upper bound on the ordinary relative entropy in terms of the trace norm distance,
we can find an upper bound on the telescopic relative entropy. This bound has a very simple form, but is nevertheless
the strongest one possible.

\begin{theorem}\label{th:1}
With $\tau=a\rho+(1-a)\sigma$,
\be
S(\rho||\tau) \le -\log(a)\,T(\rho,\sigma).\label{eq:ST1}
\ee
\end{theorem}

This theorem immediately gives our first important relation for the TRE.
\begin{corollary}
For any $a\in(0,1)$,
\be
S_a(\rho||\sigma) \le T(\rho,\sigma).
\ee
\end{corollary}
Equality can be obtained for any value of $t=T(\rho,\sigma)$ in dimension 3 and higher by choosing
$\rho=\diag(t,0,1-t)$ and $\sigma=\diag(0,t,1-t)$.

\bigskip

A second and unsuspected corollary is a strengthening of a very well-known inequality
(see, e.g.\ \cite{petzbook}, Th.\ 3.7)
for the entropy of an ensemble of two states:
for any two states $\rho,\sigma$ and $(p,1-p)$ a probability distribution,
\be\label{eq:sub1}
S(p\rho+(1-p)\sigma) \le p S(\rho) + (1-p) S(\sigma) + h(p),
\ee
where $h(p)=-p\log p-(1-p)\log(1-p)$ is the binary Shannon entropy.
This inequality is equivalent to \textit{subadditivity of the von Neumann entropy}
(w.r.t.\ ordinary addition) for positive (non-normalised) operators:
for any $A,B\ge0$
\be\label{eq:sub2}
S(A+B) \le S(A)+S(B).
\ee
Indeed, substituting $A=p\rho$ and $B=(1-p)\sigma$ yields (\ref{eq:sub1}).

The quantity $S(p\rho+(1-p)\sigma) - (p S(\rho) + (1-p) S(\sigma))$ is known as the
\textit{Holevo quantity} $\chi(\cE)$ for the ensemble $\cE=\{(p,\rho),(1-p,\sigma)\}$ (of cardinality 2).
The bound says that $\chi(\cE)\le h(p)$.
Using Theorem \ref{th:1}, we get a sharper bound:
\begin{corollary}
For any ensemble $\cE=\{(p,\rho),(1-p,\sigma)\}$ of cardinality 2,
\be
\chi(\cE) \le h(p) \,\, T(\rho,\sigma).
\ee
\end{corollary}
\textbf{Proof.}
Let $\tau=p\rho+(1-p)\sigma$.
Notice that $S(\tau) - (p S(\rho) + (1-p) S(\sigma))$
is equal to $p S(\rho||\tau)+(1-p) S(\sigma||\tau)$.
Applying inequality (\ref{eq:ST1}) to both terms gives
$-p\log(p) \,T(\rho,\sigma) -(1-p)\log(1-p) \,T(\rho,\sigma)$ as an upper bound.
\qed

\bigskip

\noindent\textbf{Question.}
As inequality (\ref{eq:sub1}) immediately generalises to ensembles of any cardinality (\cite{MI}, section 11.3.6),
namely, $\chi(\cE)\le H(p)$ (where $H(p)$ is the Shannon entropy of the probability distribution of $\cE$),
it is fair to ask for a similar generalisation of the Corollary.

In \cite{roga}, related upper bounds were studied. For cardinality 2, a bound was found in terms of the probability $p$
and the Uhlmann fidelity between $\rho$ and $\sigma$,
$F=||\sqrt{\rho}\sqrt{\sigma}||_1$. For cardinality 3, a generalisation was conjectured in \cite{roga2}. For general
cardinalities a bound was proven that is sharper than $H(p)$ and is expressed in terms of the so-called exchange entropy
\cite{roga}.

\bigskip

We now present the proof of Theorem \ref{th:1}.
It relies on the properties of the Fr\'echet derivative of the matrix logarithm given in Section \ref{sec:pre}.

\noindent\textbf{Proof of Theorem \ref{th:1}.}\\
Let $\rho$ and $\sigma$ be two given states, and $\tau=a\rho+(1-a)\sigma$.
Define $s=(1-a)/a$, which is a non-negative number. Thus
$\tau=a(\rho+s\sigma)$. W.l.o.g.\ we will assume that $\rho+s\sigma$ is full rank.

Let $\Delta:=\rho-\sigma$, $t:=T(\rho,\sigma)=||\Delta||_1/2$ and $\omega:=\Delta/t$. Obviously, $\omega$ has trace 0 and trace norm 2.
Let its Jordan decomposition be $\omega=\omega_+ - \omega_-$. Thus $\omega\le\omega_+$ and $\trace\omega_+ = \trace\omega_- = 1$.

Now consider the expression $s\trace\omega\cT_{\rho+s\sigma}(\sigma)$.
Since $\cT_{\rho+s\sigma}(\sigma)\ge0$, and $\omega\le\omega_+$, we have
\beas
s\trace\omega\cT_{\rho+s\sigma}(\sigma) &=& \trace\omega\cT_{\rho+s\sigma}(s\sigma) \\
&\le& \trace\omega_+\cT_{\rho+s\sigma}(s\sigma)\\
&\le&\trace\omega_+\cT_{\rho+s\sigma}(\rho+s\sigma)\\
&=&\trace\omega_+\id \\
&=& 1.
\eeas
Then, noting that $\rho=\sigma-t\omega$,
\beas
(1+s)\trace\rho\cT_{\rho+s\sigma}(\sigma)
&=& \trace(\rho+s\rho)\cT_{\rho+s\sigma}(\sigma) \\
&=& \trace(\rho+s\sigma-st\omega)\cT_{\rho+s\sigma}(\sigma) \\
&=& \trace(\rho+s\sigma)\cT_{\rho+s\sigma}(\sigma) -ts\trace\omega\cT_{\rho+s\sigma}(\sigma) \\
&=& \trace\sigma\cT_{\rho+s\sigma}(\rho+s\sigma) -ts\trace\omega\cT_{\rho+s\sigma}(\sigma) \\
&=& \trace\sigma -ts\trace\omega\cT_{\rho+s\sigma}(\sigma) \\
&\ge&1-t.
\eeas
Therefore,
$$
\trace\rho\cT_{\rho+s\sigma}(\sigma) \ge \frac{1-t}{1+s}.
$$
Integrating over $s$ from $0$ to $(1-a)/a$ then yields
$$
\trace\rho\log(\rho+(1-a)\sigma/a) - \trace\rho\log(\rho)\ge (1-t)\log(1/a),
$$
which becomes, after adding $\log a$ to both sides,
$$
\trace\rho\log(a\rho+(1-a)\sigma) - \trace\rho\log(\rho)\ge t\log(a),
$$
which is equivalent to the statement of the Theorem.
\qed

\section{Cases of maximality}
The following theorem characterises those cases when the telescopic relative
entropy achieves its maximal value of $1$.
\begin{theorem}
For any $a\in(0,1)$, $S_a(\rho||\sigma)=1$ iff $\rho\perp\sigma$.
\end{theorem}
\textbf{Proof.}
We have $S_a(\rho||\sigma)=1$ iff $\trace\rho\log(a\rho)=\trace\rho\log(a\rho+(1-a)\sigma)$
or, putting $X=a\rho$ and $Y=(1-a)\sigma$, iff $\trace X\log X = \trace X\log(X+Y)$.
Since $X,Y\ge0$, operator monotonicity of the logarithm gives $\trace X\log(X+Y) \ge \trace X\log X$.
We want to characterise the cases of equality.
One direction is obvious; if $X$ and $Y$ are orthogonal, clearly we have equality.

To prove that there are no other possibilities, assume $\trace X (\log(X + Y)-\log X)=0$.
Consider first the case $X>0$.
Define $Z=\log(X+Y) - \log X$.
Because of monotonicity of the logarithm we have $Z \ge 0$, hence
the assumption, $\trace XZ=0$, implies $Z=0$, i.e.\ $\log(X + Y)=\log X$.
As the logarithm is invertible on the set of positive operators,
this can only be true iff $Y=0$.

Now consider the general case $X\ge0$, and assume $X$ has a non-trivial kernel.
Then we can decompose the Hilbert space $\cH$ as the direct sum
$\cH=\supp X\oplus\ker X$. We have $X=X|_X\oplus 0$, with $X|_X>0$.
W.l.o.g.\ we can assume that $X+Y>0$, so that its logarithm is well-defined. By the convention to take
$\lim_{x\to 0}x\log x=0$, $\trace X\log X$ is well-defined, too, and equal to $\trace X|_X\log X|_X$.
The assumption $\trace X (\log(X + Y)-\log X)=0$ can then be written as
$\trace X|_X (\log(X + Y)|_X-\log (X|_X))=0$.
Let us therefore define $Z=\log(X + Y)|_X-\log (X|_X)$.

As can be expected, $Z\ge0$. To prove this,
put $X'=X|_X\oplus\epsilon\id$. By operator monotonicity of the logarithm,
$\log(X'+Y)-\log X'\ge0$, for all $\epsilon>0$. In particular, the compression to $\supp X$ is positive too:
$\log(X'+Y)|_X-\log (X')|_X\ge0$. Since $X'$ is defined as a direct sum of $X$ and $\epsilon\id$,
$\log (X')|_X = \log (X'|_X) = \log(X|_X)$. Since $\lim_{\epsilon\to0} X'+Y = X+Y$, we get, indeed,
$\log(X + Y)|_X-\log (X|_X)\ge0$.

The assumption reduces to $\trace X|_X \;Z=0$.
Because $X|_X>0$ and $Z\ge0$, this implies $Z=0$.

This implies $Y|_X=0$, so that, indeed, $Y$ must be orthogonal to $X$.
\qed
\section{Relative Renyi Entropies}
The relative Renyi entropies are parameterised modifications of the relative entropy given by
$$
\trace \rho^{1-p} \sigma^p,
$$
where $p$ is a real number. Here we restrict ourselves to
the case $0\le p\le1$.

Just as we have done for the relative entropy, one can define the telescopic relative Renyi entropy, even though the problem of infinite values
does not pose itself here; indeed, $\trace \rho^{1-p} \sigma^p$ is always between $0$ and $1$. 
Nevertheless, some interesting relationships occur when telescoping
the relative Renyi entropies. In particular, by exploiting the methods used in Section \ref{sec:tracenorm} we obtain a shorter and much simpler
proof of an inequality already proven in \cite{ka}.

Let us therefore consider the quantity $\trace\rho^{1-p}(a\rho+(1-a)\sigma)^p$. 
Firstly, let us determine its extremal values for fixed values of $a$.
Clearly, the maximum is still $1$, achieved when $\rho=\sigma$. The minimal value, however, is now $a^p$. This follows easily from
operator monotonicity of the fractional power $x\mapsto x^p$ when $0\le p\le 1$.
Indeed, 
\beas
\trace \rho^{1-p} (a\rho+(1-a)\sigma)^p 
&\ge& \trace \rho^{1-p} (a\rho)^p \\
&=& a^p \trace \rho^{1-p} \rho^p = a^p \trace\rho = a^p.
\eeas
Equality can be achieved for orthogonal $\rho$ and $\sigma$.

Hence, we define the telescopic relative Renyi entropies (TRRE) as follows:
\begin{definition}
\be
Q_{p,a}(\rho,\sigma) = \frac{1}{1-a^p} (1-\trace \rho^p(a\rho+(1-a)\sigma)^{1-p}).
\ee
\end{definition}
By the above, the TRRE has values between $0$ and $1$.

We now show that a sharper upper bound is given by the trace norm distance between $\rho$ and $\sigma$.
\begin{theorem}\label{th:QpaT}
\be
Q_{p,a}(\rho,\sigma) \le T(\rho,\sigma).
\ee
\end{theorem}
As a special case, we recover the bound $Q_{p,0}(\rho,\sigma) = 1-\trace \rho^p\sigma^{1-p} \le T(\rho,\sigma)$, which was
instrumental in proving optimality of the Chernoff bound in symmetric hypothesis testing \cite{ka}.

Just as we did for the operator logarithm, we can define a linear map based on the Fr\'echet derivative of the fractional power function $x^p$, via
$$
\frac{d}{dt}\Bigg|_{t=0} (A+t\Delta)^p =: \cT_{A;p}(\Delta).
$$
Since $x\mapsto x^p$ is a non-negative operator monotone function for $0\le p\le 1$, the fractional power of
a positive operator $A$ can be written as the integral
$$
A^p = \int_0^\infty d\mu_p(s)\; (A+s)^{-1}A,
$$
where $d\mu_p(s)$ is a certain measure, parameterised by $p$, that is positive for $0\le p\le 1$.
Its Fr\'echet derivative is therefore given by
\beas
\frac{d}{dt}\Bigg|_{t=0} (A+t\Delta)^p
&=& \int_0^\infty d\mu_p(s)\;((A+s)^{-1}\Delta - (A+s)^{-1}\Delta(A+s)^{-1}A) \\
&=& \int_0^\infty d\mu_p(s)\;s (A+s)^{-1}\Delta(A+s)^{-1}.
\eeas
Therefore, $\cT_{A;p}$ has the integral representation
\be
\cT_{A;p}(\Delta) = \int_0^\infty d\mu_p(s)\;s (A+s)^{-1}\Delta(A+s)^{-1}.
\ee
From this representation we easily derive the following properties:
\begin{enumerate}
\item $\trace X\cT_{A;p}(Y) = \trace Y\cT_{A;p}(X)$ for any $X$ and $Y$;
\item the map $\cT_{A;p}$ preserves the positive definite ordering; 
\item in particular, $\cT_{A;p}(B)$ is positive for positive $B$;
\item for $0<p<1$, $\cT_{A;p}(A^{1-p}) =  p\{A\}$.
\end{enumerate}
The last property follows from
\beas
\cT_{A;p}(A^{1-p}) &=& \frac{d}{dt}\Bigg|_{t=0} (A+tA^{1-p})^p \\
&=& pA^{p-1} A^{1-p} = p\{A\}.
\eeas
Using these properties, we can easily prove the theorem.

\bigskip

\textbf{Proof of Theorem \ref{th:QpaT}.}
Let $\Delta=\rho-\sigma$, and $t=T(\rho,\sigma)$ then $\Delta$ has Jordan decomposition $\Delta=t\omega_+ - t\omega_-$,
where $\omega_+$ and $\omega_-$ are orthogonal density operators.
Then
\beas
\trace(a\rho)^{1-p} \cT_{a\rho+(1-a)\sigma;p}(\Delta)
&\le& \trace(a\rho)^{1-p} \cT_{a\rho+(1-a)\sigma;p}(t\omega_+) \\
&\le& \trace(a\rho+(1-a)\sigma)^{1-p} \cT_{a\rho+(1-a)\sigma;p}(t\omega_+) \\
&=& \trace t\omega_+ \cT_{a\rho+(1-a)\sigma;p}((a\rho+(1-a)\sigma)^{1-p}) \\
&=& \trace t\omega_+ p \{a\rho+(1-a)\sigma\} \\
&\le& pt.
\eeas
In the first line we used the fact that $\Delta\le t\omega_+$ and property 2;
in the second line we used operator monotonicity of $x^{1-p}$ and property 3;
in the third line we used property 1, and in the fourth property 4.
In the last line we used the fact that $\trace XY\le1$ when $X$ is a density operator and $Y$ is a projector.

Exploiting the inequality just obtained in the last of the following integrals, we get
\beas
1-\trace \rho^p(a\rho+(1-a)\sigma)^{1-p} &=&
\trace\rho^{1-p}(\rho^p -(a\rho+(1-a)\sigma)^p) \\
&=& \int_a^1 da \; \frac{d}{da} \trace\rho^{1-p}(a\rho+(1-a)\sigma)^p \\
&=& \int_a^1 da \; \trace\rho^{1-p}\frac{d}{da}(a\rho+(1-a)\sigma)^p \\
&=& \int_a^1 da \; \trace\rho^{1-p}\cT_{a\rho+(1-a)\sigma;p}(\rho-\sigma) \\
&\le& \int_a^1 da \; a^{p-1} pt \\
&=& (1-a^p)t,
\eeas
which is equivalent to the statement of the theorem.
\qed

\section{Future work}
In forthcoming papers we will explore further properties of the telescopic relative entropy.
One other problem with the ordinary relative entropy is the absence of a triangle inequality, in the
sense that no useful upper bound exists on the difference $S(\rho||\tau_1) - S(\rho||\tau_2)$. Indeed,
this difference can be infinite. It turns out that such a bound does exist for the telescopic relative entropy.
Together with an upper bound on the difference $S(\rho_1||\tau)-S(\rho_2||\tau)$ it will be presented and proven in \cite{TRE2}.

We will also study an interesting connection with Hamiltonian reconstruction. There is some evidence
that the difference $S_a(\rho||\tau_1) - S_a(\rho||\tau_2)$ might provide non-trivial lower bounds
on the time needed for state $\tau_1$ to evolve unitarily into state $\tau_2$ under the influence
of a Hamiltonian with bounded energy.

%
\section*{Acknowledgments}
The main part of this work was done at the Institut Mittag-Leffler, Djurs\-holm (Sweden),
during an extended stay at its Fall 2010 Semester on Quantum Information Theory.



\begin{thebibliography}{000}
\bibitem{ka} K.M.R.~Audenaert, M.~Nussbaum, A.~Szko\l a and F.~Verstraete, {Commun.~Math.~Phys.} {\bf 279}, 251--283 (2008).
\bibitem{TRE2} K.M.R.\ Audenaert, ``Telescopic Relative Entropy -- II: Triangle inequalities'',
arxiv:1102:3041 (2011).
\bibitem{nila} N.~Datta, ``Min- and Max-Relative Entropies and a New Entanglement Monotone,''
IEEE Trans. Information Theory \textbf{55}, 2816--2826 (2009).
\bibitem{roga2} M.\ Fannes, F.\ de Melo, W.\ Roga and K.\ {\.Z}yczkowski, ``Matrices of fidelities for ensembles of
quantum states and the Holevo quantity'', arXiv:1104.2271 (2011).
\bibitem{lendi} K.~Lendi, F.~Farhadmotamed and A.J.~van~Wonderen,
``Regularization of quantum relative entropy in finite dimensions and application to entropy production'',
J.\ Stat.\ Phys.\ \textbf{92}(5/6), 1115--1135 (1998).
\bibitem{MI} M.A.~Nielsen and I.L.~Chuang, \textit{Quantum Computation and Quantum Information},
Cambridge University Press (2000).
\bibitem{ohya_petz} M.~Ohya and D.~Petz, ``Quantum entropy and its use'', Springer (1993).
\bibitem{petzbook} D.~Petz, \textit{Quantum Information Theory and Quantum Statistics},
Springer-Verlag, Berlin (2008).
\bibitem{renner} R.~Renner, ``Security of quantum key distribution,'' PhD
thesis, ETH Zurich, arXiv:quant-ph/0512258 (2005).
\bibitem{roga} W.\ Roga, M.\ Fannes and K.\ {\.Z}yczkowski, Phys.\ Rev.\ Lett.\ \textbf{105}, 040505 (2010).
\end{thebibliography}
\end{document}